\documentclass[pra,floatfix,showpacs,superscriptaddress]{revtex4}
\usepackage{amssymb}
\usepackage{amsmath}
\usepackage{graphicx}
\usepackage{float}
\usepackage{dcolumn}
\usepackage{bm}

\begin{document}

\title{Quantum state transmission via a spin ladder as a robust data bus }
\author{Y. Li}
\affiliation{Department of Physics, Nankai University, Tianjin
300071, China}
\author{T. Shi}
\affiliation{Department of Physics, Nankai University, Tianjin
300071, China}
\author{Z. Song}
\affiliation{Department of Physics, Nankai University, Tianjin
300071, China}
\author{C. P. Sun}\email{suncp@itp.ac.cn; http://www.itp.ac.cn/~suncp}
\affiliation{Department of Physics, Nankai University, Tianjin
300071, China}
\affiliation{Institute of Theoretical Physics, The Chinese
Academy of Science, Beijing, 100080, China}


\begin{abstract}
We explore the physical mechanism to coherently transfer the quantum
information of spin by connecting two spins to an isotropic
antiferromagnetic spin ladder system as data bus. Due to a large spin gap
existing in such a perfect medium, the effective Hamiltonian of the two
connected spins can be archived as that of Heisenberg type, which possesses
a ground state with maximal entanglement. We show that the effective
coupling strength is inversely proportional to the distance of the two spins
and thus the quantum information can be transferred between the two spins
separated by a longer distance, i.e. the characteristic time of quantum
state transferring linearly depends on the distance.
%
%
\end{abstract}

\maketitle

Transferring a quantum state from a quantum bit to another is not only the
central task in the quantum communication, but also is often required in
scalable quantum computing based on the quantum network \cite{q-inf}. In the
latter, one should connect different quantum predeceasing units  in
different locations with a medium called data bus. The typical examples of
quantum state transfer is the quantum storage based on various physical
systems\cite{Hau,Lukin}, such as the quasi-spin wave excitations \cite{s-prl}%
. For the solid state based quantum computing at the large-scale,
it is very crucial to have a solid system serving as such quantum
data bus, which can provide us with a quantum channel for quantum
communication \cite{q-solid}. Most recently the simple spin chain,
a typical solid state system, is considered as a coherent data bus
\cite{Bose,Subra,Matt}. The quantum transmission of state is
achieved by placing two spins at the two ends of the chain. These
schemes may admit an efficient state transfer of any quantum state
in a fixed period of time of the state evolution, but the crucial
problem is the dependence of transferring efficiency on
communication distance. In usual the  efficiency  is inversely
proportional to square or higher order power of the distance of
the two spins and thus such quantum state transmission can only
works efficiently in a much shorter distance.

The aim of this letter is to solve this short-distant transfer problem by
replacing the simple spin chain with an isotropic antiferromagnetic spin
ladder. Because the this kind of spin ladder possesses a finite spin gap, an
effective Heisenberg interaction can be induced in the stable ground state
channel to achieve the maximally entangled states that implement a more fast
quantum states transfer of two spin qubits attached to this spin ladder
system. Actually, when the spin gap is sufficiently large comparing to the
coupling strength between two spin qubits and the spin ladder, the
perturbation method can be performed. Analytical and numerical results show
that the spin ladder system is a perfect medium through which the
interaction between two distant spins can be mapped to an approximate
Heisenberg type coupling with a coupling constant inversely proportional to
the distance between the two separated spins.

It is well known that there are two ways to transfer quantum information:
one can first use the channel to share entanglement with separated Alice and
Bob and then use this entanglement for teleportation\cite{Qtele}, or
directly transmit a state through a quantum data bus. For the latter it
seems that the long distance entanglement is not necessary to interface
different kinds of physical systems, but and it will be showed in this
letter that there hides an effective entanglement intrinsically. In this
senses a quantum state transmission can be generally understood through such
quantum entanglement.

We sketch our idea with the model illustrated in Fig.1 The whole quantum
system we consider here consists of two qubits (A and B) and a $2\times N$%
-site two-leg spin ladder. In practice, this system can be realized by the
engineered array of quantum dots \cite{QD array}. The total Hamiltonian

\begin{equation}
H=H_{M}+H_{q}  \label{1}
\end{equation}

\begin{figure}[h]
\hspace{24pt}\includegraphics[width=10cm,height=15cm]{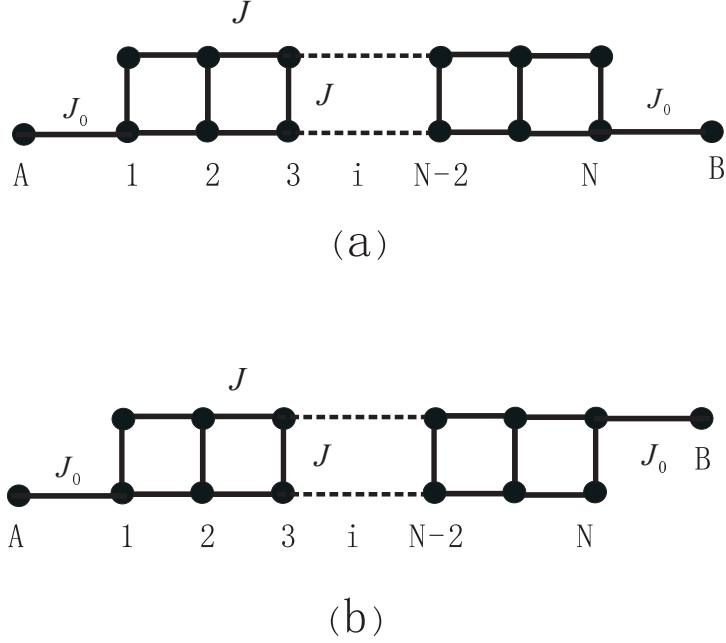} \vspace*{%
-5.0cm}
\caption{Two qubits $A$ and $B$ connect to a $2\times N$-site spin ladder.
The ground state of $H$ with a-type connection (Fig.1(a)) is singlet
(triplet) when $N$ is even (odd), while for b-type connection (Fig.1(b)),
one should have opposite result. }
\end{figure}
contains two parts, the medium Hamiltonian
\begin{equation}
H_{M}=J\sum_{\left\langle ij\right\rangle \perp }\overrightarrow{S}_{i}\cdot
\overrightarrow{S}_{j}+J\sum_{\left\langle ij\right\rangle \parallel }%
\overrightarrow{S}_{i}\cdot \overrightarrow{S}_{j}  \label{2}
\end{equation}%
describing the spin-1/2 Heisenberg spin ladder consisting of two coupled
chains and the coupling Hamiltonian
\begin{equation}
H_{q}=J_{0}\overrightarrow{S}_{A}\cdot \overrightarrow{S}_{L}+J_{0}%
\overrightarrow{S}_{B}\cdot \overrightarrow{S}_{R}  \label{3}
\end{equation}%
describing the connections between qubits $A$, $B$ and the ladder. In the
term $H_{M},$ $i$ denotes a lattice site on which one electron sits, $%
\left\langle ij\right\rangle $ $\perp $ denotes nearest neighbor sites on
the same rung, $\left\langle ij\right\rangle $ $\parallel $ denotes nearest
neighbors on either leg of the ladder. In term $H_{q}$, $L$ and $R$ denote
the sites connecting to the qubits $A$ and $B$ at the ends of the ladder.
There are two types of the connection between $\overrightarrow{S}_{A}(%
\overrightarrow{S}_{B})$ and the ladder, which are illustrated in Fig.1.
According to the Lieb's theorem \cite{Lieb}, the spin of the ground state of
$H$ with the connection of type a is zero (one) when $N$ is even (odd),
while for the connection of type b, one should have an opposite result. For
the two-leg spin ladder $H_{M},$ analytical analysis and numerical results
have shown that the ground state and the first excited state of the spin
ladder have spin $0$ and $1$ respectively \cite{Dagotto,White}. It is also
shown that there exists a finite spin gap
\begin{equation}
\triangle =E_{1}^{M}-E_{g}^{M}\sim J/2.  \label{4}
\end{equation}%
between the ground state and the first excited state (see the Fig.2). This
fact has been verified by experiments \cite{Dagotto} and is very crucial for
our present investigation.

\begin{figure}[h]
\hspace{24pt}\includegraphics[width=10cm,height=15cm]{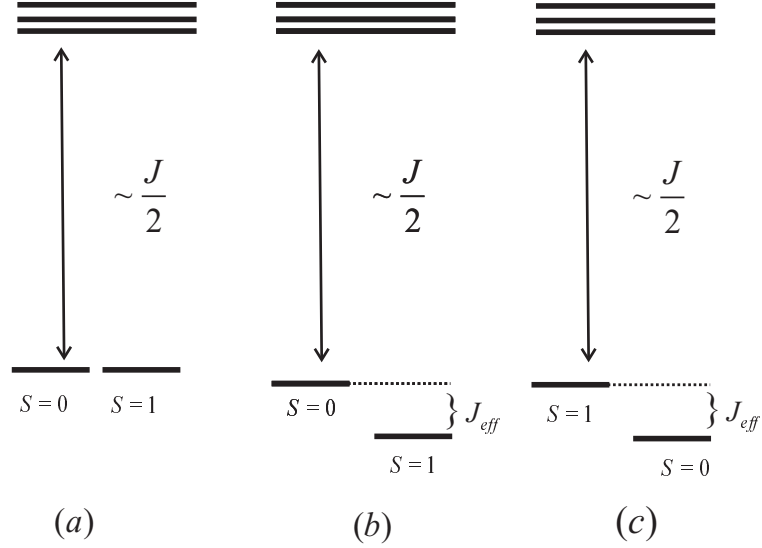} \vspace*{%
-5.0cm}
\caption{Schematic illustration of the energy levels of the system. (a) When
the connections between two qubits and the medium switch off ($J_{0}=0$) the
ground states are degenerate. (b), (c) When $J_{0}$ switches on, the ground
state(s) and the first excited state(s) are either singlet or triplet. This
is approximately equivalent to that of two coupled spins. }
\end{figure}

Thus, it can be concluded that the medium can be robustly frozen its ground
state to induced the effective Hamiltonian
\begin{equation}
H_{eff}=J_{eff}\overrightarrow{S}_{A}\cdot \overrightarrow{S}_{B}  \label{5}
\end{equation}%
between the two end qubits. With the effective coupling constant $J_{eff}$
to be calculated in the following, this Hamiltonian depicts the direct
exchange\ coupling between two separated qubits. As the famous Bell states, $%
H_{eff}$ has singlets and triplets eigenstates $\left\vert j,m\right\rangle
_{AB}:$ $\left\vert 0,0\right\rangle =$ $\frac{1}{\sqrt{2}}\left( \left\vert
\uparrow \right\rangle _{A}\left\vert \downarrow \right\rangle
_{B}-\left\vert \downarrow \right\rangle _{A}\left\vert \uparrow
\right\rangle _{B}\right) $ and $\left\vert 1,1\right\rangle =\left\vert
\uparrow \right\rangle _{A}\left\vert \uparrow \right\rangle _{B}$ ,$%
\left\vert 1,-1\right\rangle =\left\vert \downarrow \right\rangle
_{A}\left\vert \downarrow \right\rangle _{B},\left\vert 1,0\right\rangle $ $=%
\frac{1}{\sqrt{2}}\left( \left\vert \uparrow \right\rangle _{A}\left\vert
\downarrow \right\rangle _{B}+\left\vert \downarrow \right\rangle
_{A}\left\vert \uparrow \right\rangle _{B}\right) ,$which can be\ used as
the channel to share entanglement for a perfect quantum communication in a
longer distance.

The above central conclusion can be proved both with the analytical \ and
numerical methods as follows. To deduce the above effective Hamiltonian we
utilize the Fr\H{o}hlich transformation, whose original approach was used
successfully for the superconductivity BCS theory. As a second order
perturbation, the effective Hamiltonian $H_{\mathrm{eff}}\cong H_{\mathrm{M}%
}+\frac{1}{2}\left[ H_{\mathrm{q}},S\right] $ can be achieved approximately
by a unitary operator $U=\exp \{-S\}$ where anti-Hermitian operator $S$,
obeys $H_{\mathrm{q}}+[H_{\mathrm{M}},S]=0.$Let $\left\vert m\right\rangle $
and $E_{m}$ are the eigenvectors and eigenvalues of $H_{M}=$ $H(J_{0}=0)$
respectively.

From the explicit expressions for the elements $S_{mn}$ =\ $\left( H_{%
\mathrm{q}}\right) _{\mathrm{mn}}/(E_{\mathrm{m}}-E_{\mathrm{n}}),(m\neq
n),S_{mm}=0$, the matrix elements of effective Hamiltonian can be achieved
approximately as
\begin{equation}
\left\langle n\right\vert H_{\mathrm{eff}}\left\vert m\right\rangle \cong E_{%
\mathrm{m}}\delta _{mn}+\underset{\mathrm{k}\neq \mathrm{m}}{\sum }\frac{%
\left( H_{\mathrm{q}}\right) _{\mathrm{nk}}\left( H_{\mathrm{q}}\right) _{%
\mathrm{km}}}{2(E_{\mathrm{k}}-E_{\mathrm{m}})}-\underset{\mathrm{k}\neq
\mathrm{n}}{\sum }\frac{\left( H_{\mathrm{q}}\right) _{\mathrm{nk}}\left( H_{%
\mathrm{q}}\right) _{\mathrm{km}}}{2(E_{\mathrm{n}}-E_{\mathrm{k}})}.
\tag{6}
\end{equation}

We use $|\psi _{g}\rangle _{M}$ ($|\psi _{\alpha }\rangle _{M}$) and $E_{g}$
($E_{\alpha }$) to denote ground (excited) states of $H_{\mathrm{M}}$ and
the corresponding eigen-values. The zero order eigenstates $\left\vert
m\right\rangle $ can then be written as in a joint way
\begin{equation}
\left\vert j,m\right\rangle _{g}=\left\vert j,m\right\rangle _{AB}\otimes
|\psi _{g}\rangle _{M},\left\vert \psi _{\alpha }^{jm}(s^{z})\right\rangle
=\left\vert j,m\right\rangle _{AB}\otimes |\psi _{\alpha }\rangle _{M}
\tag{7}
\end{equation}%
Here, we have considered that z-component $%
S^{z}=S_{M}^{z}+S_{A}^{z}+S_{B}^{z}$ of total spin \ is conserved with
respect to the connection Hamiltonian $H_{q}$. Since $S_{M}^{z}$ and $%
S_{M}^{2}$ conserves with respect to $H_{M}$ we can label $|\psi _{g}\rangle
_{M}$ as $|\psi _{g}(s_{M},s_{M}^{z},)\rangle _{M}$ and then $%
s^{z}=m+s_{M}^{z}$ can characterize the non-coupling spin state $\left\vert
\psi _{\alpha }^{jm}(s^{z})\right\rangle .$

When the connections between two qubits and the medium switch off, i.e., $%
J_{0}=0,$ the degenerate ground states of $H$ are just $\left\vert
j,m\right\rangle _{g}$ with the degenerate energy $E_{g}$ and spin $0,1$
respectively, which is illustrated in Fig.2 (a)$.$ When the connections
between the two qubits and the medium switch on, the degenerate states with
spin $0,1$ should split as illustrated in Fig.2 (b) and (c). In the case
with $J_{0}\ll J$ at lower temperature $kT<J/2$ , the medium can be frozen
in its ground state and then we have the effective Hamiltonian

\begin{eqnarray}
H_{\mathrm{eff}} &\cong &\sum_{j^{\prime },m^{\prime },j,m,s^{z}}\frac{%
|_{g}\left\langle j,m\right\vert H_{\mathrm{q}}\left\vert \psi _{\alpha
}^{j^{\prime }m^{\prime }}(s^{z})\right\rangle |^{2}}{E_{g}-E_{\mathrm{%
\alpha }}}\left\vert j,m\right\rangle _{gg}\left\langle j,m\right\vert
\label{8} \\
&=&J_{eff}.\mathit{Diag}.(\frac{1}{4},\frac{1}{4},\frac{1}{4},-\frac{3}{4}%
)+\varepsilon   \notag
\end{eqnarray}

where%
\begin{eqnarray}
J_{eff} &=&\sum\limits_{\alpha }\frac{J_{0}^{2}[L(\alpha )R^{\ast }(\alpha
)+R(\alpha )L^{\ast }(\alpha )]}{E_{g}-E_{\alpha }},  \label{10} \\
\varepsilon  &=&\sum\limits_{\alpha }\frac{3J_{0}^{2}\left[ \left\vert
L(\alpha )\right\vert ^{2}+\left\vert R(\alpha )\right\vert ^{2}\right] }{%
4\left( E_{g}-E_{\alpha }\right) }.  \notag
\end{eqnarray}%
This just proves the above effective Heisenberg Hamiltonian (5). Here, the
matrix elements of interaction $K(\alpha )=_{M}\left\langle \psi
_{g}\right\vert S_{K}^{z}\left\vert \psi _{\alpha }\left( 1,0\right)
\right\rangle _{M}$ ($K=S,L)$ can be calculated only for the variables of
data bus medium. We also remark that, because $S^{z}$ and $S^{2}$ are
conserved for $H_{q},$ off-diagonal elements in the above effective
Hamiltonian vanish.

In temporal summary, we have shown that at lower temperature $kT<J/2,$ $H$
can be mapped to the effective Hamiltonian (5), which semmingly depicts the
direct exchange\ coupling between two separated qubits. Notice that the
coupling strength has the form $J_{eff}\sim g(L)J_{0}^{2}/J,$, where $g(L)$
is a function of $L=N+1$, the distance between the two qubits we concerned.
Here we take the $N=2$ case as an example. According to Eq.(9) one can get $%
J_{eff}=-\frac{1}{4}J_{0}^{2}/J$ and $\frac{1}{3}J_{0}^{2}/J$ when $A$ and $B
$ connect the plaquette diagonally and adjacently, respectively. This result
is in agreement to the theorem \cite{Lieb} about the ground state and the
numerical result when $J_{0}\gg J$. In general case, the behavior $g(L)$ vs $%
L$ is very crucial for quantum information since $L/\left\vert
J_{eff}\right\vert $ determines the characteristic time of quantum state
transfer between the two qubits $A$ and $B$. In order to investigate the
profile of $g(L),$ a numerical calculation is performed for the systems $%
L=4,5,6,7,8,$ and $10$, with $J=10,20,40,$ and $J_{0}=1$. The spin
gap between the ground state(s) and first excite state(s) are
calculated, which corresponds to the magnitude of $J_{eff}$. The
numerical result is plotted in Fig.3 , which indicates that
$J_{eff}\sim 1/(LJ)$. It implies that the characteristic time of
quantum state transfer linearly depends on the distance and then
guarantees the possibility to realize the entanglement of two
separated qubits in practice.

In order to verify the validity of the effective Hamiltonian $H_{eff}$, we
need to compare the the eigen states of $H_{eff}$ with those reduced states
from the eigen states of total system. In general the eigenstates of $H$ can
be written formally as
\begin{equation}
\left\vert \psi \right\rangle =\sum_{jm}c_{jm}\left\vert j,m\right\rangle
_{AB}\otimes |\beta _{jm}\rangle _{M}
\end{equation}%
where \{$|\beta _{jm}\rangle _{M}$ \} is a set of vectors of the data bus,
which is not necessarily orthogonal. Then we have the condition $%
\sum_{jm}|c_{jm}|_{M}^{2}\langle \beta _{jm}|\beta _{jm}\rangle
_{M}=1$for normalization of $\left\vert \psi \right\rangle $. In
this sense the practical description of the A-B subsystem of two
quits can be only given by the reduced density matrix

\begin{figure}[h]
\hspace{24pt}\includegraphics[width=10cm,height=15cm]{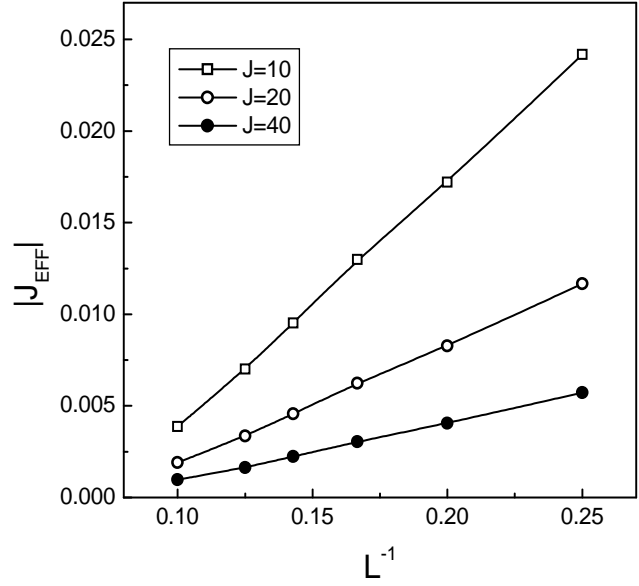} \vspace*{%
-5.0cm}
\caption{The spin gaps obtained by numerical method for the systems $%
L=4,5,6,7,8,$ and $10$, with $J=10,20,40,$ and $J_{0}=1$ are
potted, which
is corresponding to the magnitude of $J_{eff}$. It indicates that $%
J_{eff}\sim 1/(LJ)$. }
\end{figure}

\begin{eqnarray}
\rho _{AB} &=&Tr_{M}(\left\vert \psi \right\rangle \left\langle \psi
\right\vert )=\sum_{jm}|c_{jm}|^{2}\left\vert j,m\right\rangle _{AB}\langle
j,m| \\
&&+\sum_{j^{\prime }m^{\prime }\neq jm}c_{j^{\prime }m^{\prime }}^{\ast
}c_{jmM}\langle \beta _{j^{\prime }m^{\prime }}|\beta _{jm}\rangle
_{M}\left\vert j,m\right\rangle _{AB}\langle j^{\prime },m^{\prime }|  \notag
\end{eqnarray}%
where $Tr_{M}$ means the trace over the variables of the medium. By a
straightforward calculation we have
\begin{eqnarray}
\left\vert c_{11}\right\vert ^{2} &=&\left\vert c_{1-1}\right\vert
^{2}=\left\langle \psi \right\vert \left( \frac{1}{4}+S_{A}^{z}\cdot
S_{B}^{z}\right) \left\vert \psi \right\rangle ,  \notag \\
\left\vert c_{00}\right\vert ^{2} &=&\left\langle \psi \right\vert \left(
\frac{1}{4}-\overrightarrow{S}_{A}\cdot \overrightarrow{S}_{B}\right)
\left\vert \psi \right\rangle , \\
\left\vert c_{10}\right\vert ^{2} &=&1-2\left\vert c_{11}\right\vert
^{2}-\left\vert c_{00}\right\vert ^{2}.  \notag
\end{eqnarray}

Now we need a criteria to judge how close the practical reduced eigenstate
by the above reduced density matrix (11) to the pure state for the effective
two sites coupling $H_{eff}$. As we noticed, it has the singlet and triplet
eigenstates $\left\vert j,m\right\rangle _{AB}$ in the subspace spanned by $%
\left\vert 0,0\right\rangle _{AB}$ with $S^{z}=S_{A}^{z}+S_{B}^{z}=0$, we
have $\left\vert c_{11}\right\vert ^{2}=\left\vert c_{10}\right\vert
^{2}=\left\vert c_{1-1}\right\vert ^{2}=0,$ $\left\vert c_{00}\right\vert
^{2}=1;$ for triplet eigen state $\left\vert 1,0\right\rangle _{AB}$, we
have $\left\vert c_{11}\right\vert ^{2}=\left\vert c_{1-1}\right\vert
^{2}=\left\vert c_{00}\right\vert ^{2}=0,$ $\left\vert c_{10}\right\vert
^{2}=1$. With the practical Hamiltonian $H,$ the values of $\left\vert
c_{jm}\right\vert ^{2},\quad i=1,2,3,4$ are numerically calculated for the
ground state $\left\vert \psi _{g}\right\rangle $ and first excited state $%
\left\vert \psi _{1}\right\rangle $ of finite system systems $L=4,5,6,7,8$
and $10$ with $J=10,20,$ and $40,$ $(J_{0}=1)$ in $S^{z}=0$ subspace, which
are listed in the Table 1(a,b,c). It shows that, at lower temperature, the
realistic interaction leads to the results about $\left\vert
c_{jm}\right\vert ^{2}$, which are very close to that described by $H_{eff},$
even if $J$ is not so large in comparison with $J_{0}.$

\begin{center}
$%
\begin{tabular}{cccccccccc}
\hline\hline
States & $\ j\ $ & $m$ & $\ \ L\ \ $ & \ \ 4\ \ \  & \ \ \ \ 5\ \ \ \  & \ \
\ 6\ \ \  & \ \ \ \ 7\ \ \ \  & \ \ \ \ 8\ \ \ \  & \ \ \ 10\ \ \  \\ \hline
&  &  & $\left\vert c_{00}\right\vert ^{2}$ & 4.2$\times $10$^{-4}$ & 5.9$%
\times $10$^{-4}$ & 7.4$\times $10$^{-4}$ & 8.7$\times $10$^{-4}$ & 9.7$%
\times $10$^{-4}$ & 1.2$\times $10$^{-3}$ \\
$\left\vert \psi _{g}\right\rangle $ & $1$ & $0$ & $\left\vert
c_{10}\right\vert ^{2}$ & 0.9952 & 0.9954 & 0.9954 & 0.9955 & 0.9956 & 0.9956
\\
&  &  & $\left\vert c_{11}\right\vert ^{2}$ & 2.2$\times $10$^{-3}$ & 2.0$%
\times $10$^{-3}$ & 1.9$\times $10$^{-3}$ & 1.8$\times $10$^{-3}$ & 1.7$%
\times $10$^{-3}$ & 1.6$\times $10$^{-3}$ \\
&  &  & $\left\vert c_{1-1}\right\vert ^{2}$ & 2.2$\times $10$^{-3}$ & 2.0$%
\times $10$^{-3}$ & 1.9$\times $10$^{-3}$ & 1.8$\times $10$^{-3}$ & 1.7$%
\times $10$^{-3}$ & 1.6$\times $10$^{-3}$ \\ \hline
&  &  & $\left\vert c_{00}\right\vert ^{2}$ & 0.9989 & 0.9984 & 0.9979 &
0.9975 & 0.9971 & 0.9966 \\
$\left\vert \psi _{1}\right\rangle $ & $0$ & $0$ & $\left\vert
c_{10}\right\vert ^{2}$ & 3.7$\times $10$^{-4}$ & 5.2$\times $10$^{-4}$ & 7.0%
$\times $10$^{-4}$ & 8.4$\times $10$^{-4}$ & 1.0$\times $10$^{-3}$ & 1.2$%
\times $10$^{-3}$ \\
&  &  & $\left\vert c_{11}\right\vert ^{2}$ & 3.7$\times $10$^{-4}$ & 5.4$%
\times $10$^{-4}$ & 7.0$\times $10$^{-4}$ & 8.3$\times $10$^{-4}$ & 9.3$%
\times $10$^{-4}$ & 1.1$\times $10$^{-3}$ \\
&  &  & $\left\vert c_{1-1}\right\vert ^{2}$ & 3.7$\times $10$^{-4}$ & 5.4$%
\times $10$^{-4}$ & 7.0$\times $10$^{-4}$ & 8.3$\times $10$^{-4}$ & 9.3$%
\times $10$^{-4}$ & 1.1$\times $10$^{-3}$ \\ \hline
\end{tabular}%
$ \vspace*{1cm}

Table 1 (a)

$%
\begin{tabular}{llllllllll}
\hline\hline
States & $\ j\ $ & $m$ & $\ \ L\ \ $ & \ \ \ \ 4\ \ \ \  & \ \ \ \ 5\ \ \ \
& \ \ \ 6\ \ \  & \ \ \ \ 7\ \ \ \  & \ \ \ \ 8\ \ \ \  & \ \ \ 10\ \ \  \\
\hline
&  &  & $\left\vert c_{00}\right\vert ^{2}$ & 9.7$\times $10$^{-5}$ & 1.4$%
\times $10$^{-4}$ & 1.8$\times $10$^{-4}$ & 2.1$\times $10$^{-4}$ & 2.3$%
\times $10$^{-4}$ & 3.7$\times $10$^{-4}$ \\
$\left\vert \psi _{g}\right\rangle $ & $1$ & $0$ & $\left\vert
c_{10}\right\vert ^{2}$ & 0.9989 & 0.9989 & 0.9989 & 0.9989 & 0.9990 & 0.9989
\\
&  &  & $\left\vert c_{11}\right\vert ^{2}$ & 5.3$\times $10$^{-4}$ & 4.8$%
\times $10$^{-4}$ & 4.7$\times $10$^{-4}$ & 4.4$\times $10$^{-4}$ & 4.0$%
\times $10$^{-4}$ & 3.8$\times $10$^{-4}$ \\
&  &  & $\left\vert c_{1-1}\right\vert ^{2}$ & 5.3$\times $10$^{-4}$ & 4.8$%
\times $10$^{-4}$ & 4.7$\times $10$^{-4}$ & 4.4$\times $10$^{-4}$ & 4.0$%
\times $10$^{-4}$ & 3.8$\times $10$^{-4}$ \\ \hline
&  &  & $\left\vert c_{00}\right\vert ^{2}$ & 0.9997 & 0.9996 & 0.9995 &
0.9994 & 0.9993 & 0.9991 \\
$\left\vert \psi _{1}\right\rangle $ & $0$ & $0$ & $\left\vert
c_{10}\right\vert ^{2}$ & 9.1$\times $10$^{-5}$ & 1.4$\times $10$^{-4}$ & 1.7%
$\times $10$^{-4}$ & 2.0$\times $10$^{-4}$ & 2.7$\times $10$^{-4}$ & 3.7$%
\times $10$^{-4}$ \\
&  &  & $\left\vert c_{11}\right\vert ^{2}$ & 9.1$\times $10$^{-5}$ & 1.3$%
\times $10$^{-4}$ & 1.7$\times $10$^{-4}$ & 2.0$\times $10$^{-4}$ & 2.1$%
\times $10$^{-4}$ & 2.7$\times $10$^{-4}$ \\
&  &  & $\left\vert c_{1-1}\right\vert ^{2}$ & 9.1$\times $10$^{-5}$ & 1.3$%
\times $10$^{-4}$ & 1.7$\times $10$^{-4}$ & 2.0$\times $10$^{-4}$ & 2.1$%
\times $10$^{-4}$ & 2.7$\times $10$^{-4}$ \\ \hline
\end{tabular}%
$ \vspace*{1cm}

Table 1 (b)

\vspace*{1cm}

$%
\begin{tabular}{llllllllll}
\hline\hline
States & $\ j\ $ & $m$ & $\ \ L\ \ $ & \ \ \ \ 4\ \ \ \  & \ \ \ \ 5\ \ \ \
& \ \ \ 6\ \ \  & \ \ \ \ 7\ \ \ \  & \ \ \ \ 8\ \ \ \  & \ \ \ 10\ \ \  \\
\hline
&  &  & $\left\vert c_{00}\right\vert ^{2}$ & 2.3$\times $10$^{-5}$ & 3.3$%
\times $10$^{-5}$ & 4.2$\times $10$^{-5}$ & 5.0$\times $10$^{-5}$ & 5.7$%
\times $10$^{-5}$ & 1.8$\times $10$^{-4}$ \\
$\left\vert \psi _{g}\right\rangle $ & $1$ & $0$ & $\left\vert
c_{10}\right\vert ^{2}$ & 0.9997 & 0.9997 & 0.9997 & 0.9997 & 0.9998 & 0.9996
\\
&  &  & $\left\vert c_{11}\right\vert ^{2}$ & 1.3$\times $10$^{-4}$ & 1.2$%
\times $10$^{-4}$ & 1.1$\times $10$^{-4}$ & 1.1$\times $10$^{-4}$ & 8.8$%
\times $10$^{-5}$ & 9.3$\times $10$^{-5}$ \\
&  &  & $\left\vert c_{1-1}\right\vert ^{2}$ & 1.3$\times $10$^{-4}$ & 1.2$%
\times $10$^{-4}$ & 1.1$\times $10$^{-4}$ & 1.1$\times $10$^{-4}$ & 8.8$%
\times $10$^{-5}$ & 9.3$\times $10$^{-5}$ \\ \hline
&  &  & $\left\vert c_{00}\right\vert ^{2}$ & 0.9999 & 0.9999 & 0.9999 &
0.9998 & 0.9998 & 0.9997 \\
$\left\vert \psi _{1}\right\rangle $ & $0$ & $0$ & $\left\vert
c_{10}\right\vert ^{2}$ & 2.5$\times $10$^{-5}$ & 3.5$\times $10$^{-5}$ & 4.6%
$\times $10$^{-5}$ & 1.0$\times $10$^{-4}$ & 1.2$\times $10$^{-4}$ & 1.7$%
\times $10$^{-4}$ \\
&  &  & $\left\vert c_{11}\right\vert ^{2}$ & 2.3$\times $10$^{-5}$ & 3.3$%
\times $10$^{-5}$ & 4.2$\times $10$^{-5}$ & 5.0$\times $10$^{-5}$ & 4.2$%
\times $10$^{-5}$ & 6.5$\times $10$^{-5}$ \\
&  &  & $\left\vert c_{1-1}\right\vert ^{2}$ & 2.3$\times $10$^{-5}$ & 3.3$%
\times $10$^{-5}$ & 4.2$\times $10$^{-5}$ & 5.0$\times $10$^{-5}$ & 4.2$%
\times $10$^{-5}$ & 6.5$\times $10$^{-5}$ \\ \hline
\end{tabular}%
$ \vspace*{1cm}

Table 1(c)
\end{center}

\textit{Table 1. The diagonal elements of reduced density matrix, which
provide a criteria for the validity of $H_{eff},$, are calculated
numerically for the ground state and first excited state of finite system
systems $L=4,5,6,7,8$ and $10.$ The results for $J=10,20,$ and $40,(J_{0}=1)$
are listed in (a), (b), and (c) respectively. It shows that, at lower
temperature, the result based the realistic interaction is very close to
that by $H_{eff}.$}

We remark that the above tables reflect all the facts distinguishing the
difference between the results about the entanglement of two end qubit
generated by $H_{eff}$ and $H.$ Though we have ignored the considerations
for the off-diagonal terms in the reduced density matrix, the calculation of
the feudality $F(|j,m\rangle )\equiv ._{M}\langle j,m|\rho _{AB}|j,m\rangle
_{M}=|c_{jm}|^{2}$ further confirm our observation that, the effective
Heisenberg type interaction of two end qubits can approximates the realistic
Hamiltonian very well.  Then we can transfer the quantum information between
two ends of the $2\times N$-site two-leg spin ladder that can be regarded as
the channel to share entanglement with separated Alice and Bob. Physically,
this is just due to a large spin gap existing in such a perfect medium,
whose ground state can induce a maximal entanglement of the two end qubits.
We also pointed out that our analysis is applicable for other types of
medium systems as data buses, which possess a finite spin gap. Since $%
L/\left\vert J_{eff}\right\vert $ determines the characteristic time of
quantum state transfer between the two qubits, the dependence of $J_{eff}$
upon $L$ becomes important and relies on the appropriate choice of the
medium.

In conclusion, we have presented and studied in details a protocol to
achieve the entangled states and fast quantum states transfer of two spin
qubits by connecting two spins to a medium which possesses a spin gap. A
perturbation method, the Fr\H{o}hlich transformation, shows that the
interaction between the two spins can be mapped to the Heisenberg type
coupling. Numerical results show that the isotropic antiferromagnetic spin
ladder system is a perfect medium through which the interaction between two
separated spins is very close to the Heisenberg type coupling with a
coupling constant inversely proportional to the distance even if the spin
gap is not so large comparing to the couplings between the input and output
spins with the medium.

This work of SZ is supported by the Cooperation Foundation of Nankai and
Tianjin university for research of nanoscience. CPS also acknowledge the
support of the CNSF (grant No. 90203018) , the Knowledge Innovation Program
(KIP) of the Chinese Academy of Sciences, the National Fundamental Research
Program of China (No. 001GB309310).

\end{document}